\input{aipcheck}

\documentclass[final]{aipproc}

\layoutstyle{6x9}

\input{epsf}

\begin{document}

\title{Magnetorotational Instability in a Couette Flow of Plasma}

\author{Koichi Noguchi}{
  address={Applied Physics Division, Los Alamos National Laboratory, 
Los Alamos, NM 87545, USA}
}

\author{Vladimir I. Pariev}{
  address={Department of Physics and Astronomy,
University of Rochester, Rochester, NY 14627, USA
and Lebedev Physical Institute, Leninsky Prospect 53,
Moscow 119991, Russia}  
}

\begin{abstract}
All experiments, which have been proposed so far to model
the magnetorotational instability
(MRI) in the laboratory, involve a Couette flow of liquid metals in a 
rotating annulus. All liquid metals
have small magnetic Prandtl numbers, $\mbox{Pm} \sim 10^{-6}$, 
the ratio of kinematic viscosity 
to magnetic diffusivity. With plasmas both large and small
$\mbox{Pm}$ are achievable by varying the temperature 
and the density of plasma. 
Compressibility and fast rotation of the plasma
result in radial stratification of
the equilibrium plasma density. 
Evolution of perturbations
in radially stratified viscous and resistive plasma permeated by 
an axial uniform magnetic field is considered. The differential 
rotation of the plasma is induced by the 
${\bf E}\times{\bf B}$ drift in applied radial electric field. 
Global unstable eigenmodes are calculated by our newly 
developed matrix code. The plasma is shown to be MRI unstable for 
parameters easily achievable in experimental setup.
\end{abstract}

\maketitle

\newcommand{\ve}[1]{\mbox{\boldmath$ #1 $}}
\newcommand{\parti}[2]{\frac{\partial  #1 }{\partial  #2 }}
\newcommand{\calF}{{\cal{F}}}
\newcommand{\calG}{{\cal{G}}}
\renewcommand{\t}[1]{\widetilde{#1}}

\section{Introduction}\label{sec_intro}

A central problem in the theory of accretion disks in astrophysics 
is understanding the
fundamental mechanism of angular momentum transfer. 
A robust anomalous outward angular momentum transport must
operate in order for accretion to occur \cite{frank02}. 
A phenomenological theory of turbulent angular momentum transport
($\alpha$-disks) proposed by Shakura \cite{shakura72}   
was in the basis of our understanding of how accretion occurs
and still remains viable to date. The puzzle of the origin of 
turbulence in hydrodynamically stable disks was resolved when
magnetorotational instability (MRI), originally discovered
in Refs.~\cite{velikhov59} and~\cite{chandra60}, 
was applied to accretion disks by Balbus and Hawley~\cite{balbus91}.
MRI causes MHD turbulence to develop in an initially 
weakly magnetized fluid. 

Despite its importance, MRI has never 
been observed in the laboratory. 
Recently, two experiments have 
been proposed to test
MRI in a differentially rotating flow of liquid metal (Couette
flow) between two rotating cylinders \cite{ji01,noguchi02}. A great
deal of theoretical work on investigating MRI in Couette flow of 
liquid metals has been done confirming that 
MRI can be excited for magnetic Reynolds number, $\mbox{Rm}$,
exceeding a few \cite{goodman02, rudiger01, rudiger03, willis02}.
A particular attention was given to why MRI was
not observed in previous experiments with hydromagnetic Couette 
flows of liquid metals \cite{goodman02, rudiger03}. 
Because of the very small magnetic Prantdl number of metals, 
$\mbox{Pm} \sim 10^{-6}$, the rotation needs to be very fast to
achieve $\mbox{Rm}> 1$. Indeed, kinetic Reynolds number needs to exceed 
$\mbox{Re}=\mbox{Rm}/\mbox{Pm} \sim 10^6$. In previous experiments
the rotation speed was not high enough to achieve 
$\mbox{Re} \sim 10^6$. For such high Reynolds
numbers the flow in the experiment is likely to become turbulent even 
without a magnetic field for Rayleigh stable rotation profiles. 
The instability may have a nonlinear hydrodynamic nature 
\cite{richard99} or may be due to the Ekman circulation induced 
by the end plates \cite{noguchi02}. The presence of such turbulence
can affect the conditions for the excitation of MRI 
\cite{noguchi02}. 

Here we consider plasma as alternative to liquid metals to 
use in MRI experiment. By changing temperature and density of
the plasma by a few times around values $T\sim 5\,\mbox{eV}$
and $n \sim 10^{14}\,\mbox{cm}^{-3}$ one can vary $\mbox{Pm}$ 
in the range of about $10^{-2}$ to $10^{2}$. 
For velocities of plasma of the order of the thermal speed of ions
and the typical size of the apparatus of about $50\,\mbox{cm}$ 
$\mbox{Rm}$ is in the range from $10^2$ to $10^3$, 
while $\mbox{Re}$ is in the range from $5$ to $10^4$. 
These plasma parameters are readily achievable in the laboratory
\cite{wang02}. Therefore, in plasma experiment it can be relatively
easy to have high enough $\mbox{Rm}$ allowing for MRI to grow, while 
keeping $\mbox{Re}$ modest and the flow laminar. Laminar character
of the flow can allow a detailed study of the structure of the 
unstable mode and the secondary flow without intervening noise 
from turbulence. For higher $\mbox{Re} \sim 10^4$, the transition 
from laminar to turbulent flows in hydromagnetics can be investigated.
Moreover, the effects of wide variations of $\mbox{Pm}$ can be 
studied with a plasma MRI experiment.

\section{Description of the Experiment}\label{sec_setup}

\begin{figure}
\includegraphics[height=.3\textheight]{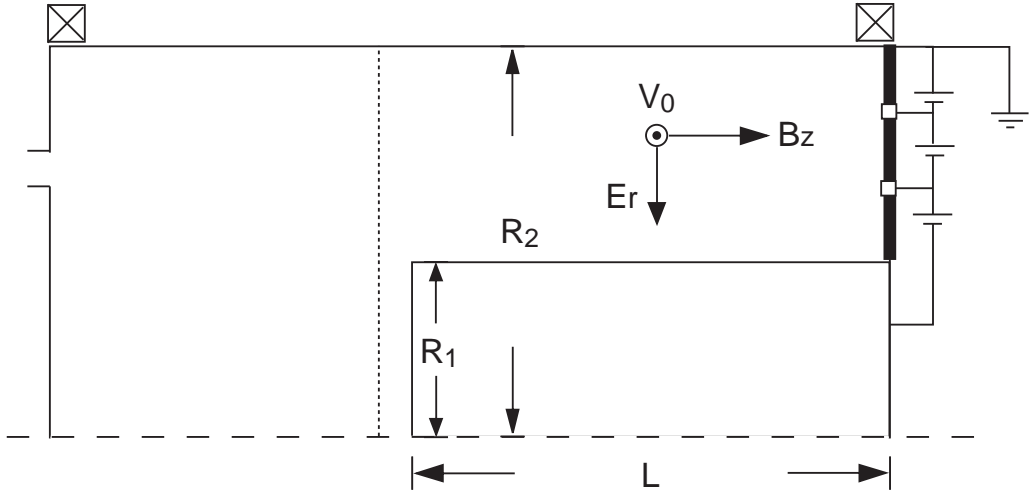}
\caption{The meridional cross section of the device. Center line of the
chamber is shown as a dashed line. Dotted line indicates the boundary 
of the working space between two cylinders (to the right).  
Plasma is injected from the left and 
rotates due to ${\bf E}\times{\bf B}$ drift. 
Axial magnetic field is produced by coils. Coaxial metal rings at the 
right end plate are charged to fractions of $\Phi_0$ ($\Phi_0<0$ as 
illustrated) to support differential rotation of plasma against the 
formation of Ekman layer.
}
\label{fig_device}
\end{figure}

The basic setup of possible plasma MRI experiment is illustrated
in Fig.~\ref{fig_device}. This experiment is now under construction
at Los Alamos National Laboratory. It can be also used for observing
laminar plasma dynamos \cite{wang02}. 
Tentative set of specifications 
and typical parameters of plasma
for that experiment is shown in Table~\ref{table1}.
The plasma is produced by the electric discharge in hydrogen and is   
injected into the space between two coaxial 
conducting cylinders. 
Mean free path for Coulomb collisions
is much smaller than the radii of the cylinders, and MHD description
of plasma is appropriate. Cylindrical coordinate system $r$, $\phi$,
$z$ is used with the axis of symmetry coinciding with the 
central axis of the cylinders.  
Axial uniform magnetic field $B_{0z}$ is produced by the coils around 
the cylinders and is preexistent of the discharge. Therefore, plasma 
is coupled to the axial magnetic field lines already at the formation and
can slide along the axial magnetic field lines filling the space 
between two cylinders. 
The outer cylinder is grounded and the voltage 
$\Phi_0$ is applied to the inner cylinder 
during and after the plasma injection, in order to maintain the rotation.
Neglecting the boundary and sheath modification of any applied
electric field, plasma rotates with ${\bf E}\times{\bf B}$ drift,
\begin{equation}
{\bf E}_0=\frac{\Phi_0}{\ln(R_2/R_1)}\frac{1}{r}{\bf e}_r
=-\frac{{\bf V}_0\times{\bf B}_0}{c},
\end{equation}
where ${\bf E}_0$ is the equilibrium electric field, ${\bf B}_0=B_{0z} 
{\bf e}_z$ is the equilibrium axial magnetic field, $R_1$ is the 
radius of the inner cylinder, $R_2$ is the radius of the outer cylinder, 
and $c$ is the speed of light.

Since plasma is highly collisional and is 
thermalized before it reaches to the 
experimental region of the chamber, we further 
make an assumption that the 
stationary state is isothermal:
\begin{equation}
\Gamma P_0/\rho_0=C_s^2=\mbox{constant},
\end{equation}
where $P_0(r)$ is the equilibrium pressure, $\rho_0(r)$ is the equilibrium
density, $C_s$ is the sound speed, and $\Gamma$ is the ratio of 
specific heats ($\Gamma=5/3$ for hydrogen). 

The Larmor radius of 
electrons in the experiment is 
also much smaller than the radii of the cylinders, but is comparable
to the Coulomb mean free path. Despite this, here we assume for 
simplicity that the conductivity of the plasma is isotropic and 
is given by non-magnetized expression. 
The coefficient of magnetic diffusivity, $\eta$, and 
dynamic viscosity coefficient $\rho\nu$ are independent of 
density and depends only on temperature. 
Then, $\eta$ and $\rho\nu$ remain constant throughout the 
volume of the plasma in the isothermal approximation considered
here. 

In the equilibrium state of the steady rotation the magnetic
field is
$B_{0z}=\mbox{constant}$ and the velocity is ${\bf V}_0=(0,r\Omega(r),0)$, 
with the angular velocity profile
\begin{equation}
\Omega(r)=-\frac{\Phi_0}{B_{0z}\ln(R_2/R_1)}\frac{c}{r^2}.\label{omega}
\end{equation}
Balance between
the centrifugal force and the gradient of pressure in the 
equilibrium state is
\begin{equation}
-r\Omega^2\rho_0=-\frac{d P_0}{dr}.\label{press_balance}
\end{equation}
Using isothermal condition, Eq.~(\ref{press_balance}) can be solved
to give the radial dependence of $\rho_0$ 
\begin{equation}
\rho_0 =\rho_0(R_1)\exp\left[\int_{R_1}^r 
\frac{r\Gamma \Omega^2}{C_s^2}dr\right]
\mbox{.}\label{rho_0}
\end{equation}
When the rotation speed $\Omega(R_1)R_1$ is comparable to the sound 
speed, the centrifugal force is significant enough to cause the 
compression of the plasma toward the outer cylinder. 

\begin{table}
\begin{tabular}{cc}
\hline
Inner Radius $r_1$ (cm)&$15$\\
Outer Radius $r_2$ (cm)&$52$\\
Length $L$ (cm)&100\\
Density n ($\mbox{cm}^{-3}$)&$1\times10^{14}$\\
Electron Temperature $T_e$ (eV)&$5$\\
Kinematic Viscosity, $\nu$ ($\mbox{cm}^2\,\mbox{s}^{-1}$)&$3\times10^6$\\
Magnetic Diffusivity, $\eta$ ($\mbox{cm}^2\,\mbox{s}^{-1}$)&$2.7\times10^5$\\
Prandtl number $P_m$&$11$\\
Sound Speed $C_s$ ($\mbox{cm}\,\mbox{s}^{-1}$)&$4\times10^6$\\
Maximum Frequency $\Omega_{1{\rm max}}$ ($\mbox{s}^{-1}$)&$1.9\times10^5$\\
\end{tabular}
\caption{Parameters of the plasma.}
\label{table1}
\end{table}

\section{The Instability in a Radially Stratified Plasma}

We consider perturbations of the equilibrium state described above:
${\bf V}={\bf V}_0+{\bf v}$, ${\bf B}={\bf B}_0+{\bf b}$,
$\rho=\rho_0+\rho_1$, $P=P_0+P_1$, 
and linearize MHD equations in small perturbations.
We idealize the problem by considering cylinders of infinite length
and take the dependence of perturbations on $t$, $\theta$, and $z$ 
as $\propto\exp[i(-\omega t+m\theta+k_z z)]$. 
If the equilibrium density $\rho_0(r)$ varies significantly 
between $R_1$ and $R_2$, general perturbations of the plasma 
are compressible. However, the phenomenon of MRI is due to the 
stretching of the magnetic field lines by the differential rotation
coupled with the action of centrifugal force. 
Boussinesq approximation neglects
the changes of the volume of the displaced parcel of plasma as it gets
quickly adjusted to the new pressure equilibrium at a new radial 
location in the stratified plasma. It allows to capture centrifugal 
force acting on a displaced parcel of plasma but excludes compressible
modes, which are not essential for the development of MRI. This simplifies
calculations substantially. Here we adopt Boussinesq approximation.  
Linearized continuity equation becomes equivalent to two equations:
\begin{eqnarray}
&& \frac{\partial \rho_1}{\partial t}+({\bf V}_0 \nabla)\rho_1 +
{\bf v}\cdot \nabla\rho_0=0 \mbox{,} \label{rho_1} \\
&& \nabla\cdot {\bf v}=0 \label{div_v}\mbox{.}
\end{eqnarray}

Other linearized MHD equations are:
\begin{eqnarray}
&& \nabla\cdot {\bf b}=0 \mbox{,}\label{lin_b1} \\
&& \frac{\partial {\bf b}}{\partial t}=\nabla\times 
({\bf v}\times {\bf B}_0) + \nabla\times ({\bf V}_0\times
{\bf b}) + \eta \nabla^2 {\bf b} \mbox{,}\label{lin_b3} \\
&& \rho_0 \frac{\partial {\bf v}}{\partial t} + \rho_0 
({\bf v}\nabla){\bf V}_0 + \rho_0({\bf V}_0\nabla){\bf v}
+\rho_1({\bf V}_0\nabla){\bf V}_0 = 
-\nabla P_1 + \rho_0 \nu_0 \nabla^2 {\bf v} \nonumber \\
&& -\frac{1}{4\pi}
\nabla(B_{0z} b_z) + \frac{1}{4\pi} ({\bf b}\nabla) {\bf B}_0
+ \frac{1}{4\pi} ({\bf B}_0 \nabla){\bf b} \mbox{,} \label{lin_b4}
\end{eqnarray}
where $\nu_0$ is the unperturbed value of kinematic viscosity. 
Note, that 
$\rho\nu=\rho_0 \nu_0$, so that the viscosity coefficient $\rho\nu$
remains unperturbed. 
The following reductions are done.
First, we note that any solution of Eq.~(\ref{lin_b3}) satisfies
Eq.~(\ref{lin_b1}) automatically. This means that out of four 
equations provided by (\ref{lin_b1}) and~(\ref{lin_b3}) together, one 
should be omitted. We choose to omit $z$-component of 
Eq.~(\ref{lin_b3}) and use 
Eq.~(\ref{lin_b1}) to express $b_z$ via $b_r$ and $b_\theta$ in
the $r$ and $\theta$ components of Eq.~(\ref{lin_b3}) as 
well as in all components of Eq.~(\ref{lin_b4}). 
We also use Eq.~(\ref{div_v}) to express $v_z$ via 
$v_r$ and $v_\theta$. 
Eq.~(\ref{rho_1}) is used to express $\rho_1$ via $v_r$
and substitute into momentum Eq.~(\ref{lin_b4}).
As for Eq.~(\ref{lin_b4}) we replace its $z$-component
by the equation obtained by taking the divergence of 
Eq.~(\ref{lin_b4}). Using Eqs.~(\ref{lin_b1}) 
and~(\ref{div_v}) the divergence of Eq.~(\ref{lin_b4}) can
be reduced to the equation with the only second order radial
derivative being the $\partial^2 \Pi/\partial r^2$,
where $\Pi=P_1+B_{0z}b_z/(4\pi)$ is the perturbation of the 
total pressure. 
In summary, we obtain five second order linear differential 
equations in $r$ for 
five variables, $\Pi$, $b_r$, $b_\theta$, $v_r$, and $v_{\theta}$:
the divergence of Eq.~(\ref{lin_b4}), $r$ and $\theta$   
components of induction Eq.~(\ref{lin_b3}), and 
$r$ and $\theta$
components of momentum Eq.~(\ref{lin_b4})

An ideal conductor boundary is a good approximation, since $\eta$ 
for metallic walls is $< 800\,\mbox{cm}^2 \,\mbox{s}^{-1}$, 
which is always much smaller than that of plasma, 
so even a thin metallic wall can be considered as a good conductor.
Boundary conditions for conducting walls are given by
$b_r=0$, $d b_\theta/d r+b_\theta/r=0$, $v_r=0$, $v_\theta=0$,
$d v_r/d r=0$ at both $r=R_1$ and $R_2$. 

We discretize the equations and boundary conditions on a uniform 
grid in rescaled variable $x=\ln (r/R_1)$ with $N=200$ grid points. 
The problem is reduced to finding values of $\omega$ such that the 
determinant of the matrix $5N\times 5N$ is zero. We calculate the 
determinant by using LU-decomposition and search for zeros by using
Newton iteration method. After an eigenvalue of $\omega$ is found,
the corresponding eigenfunction is calculated by using back 
substitution. 

\begin{figure}
\includegraphics[width=10cm]{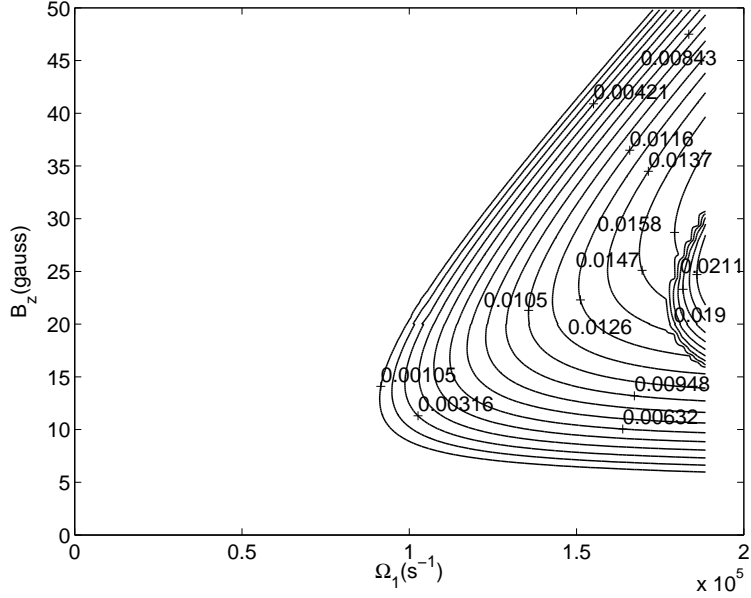}
\caption{Growth rate, $\mbox{Im}\,\omega/\Omega_1$,  
of the unstable $m=0$, $k_z=\pi/L$  
mode as a function of $\Omega_1$ and $B_{0z}$ 
}
\label{fig1}
\end{figure}

We find $m=0$ axisymmetric unstable mode in a wide range of 
parameters achievable in the experiment.
Let us focus on one case with fixed temperature and density
given in Table~\ref{table1}. Then, $C_s(T)$ and 
$\mbox{Pm}(\rho,T)$ are
also fixed. The remaining parameters, which can be adjusted in
the experiment, are $\Phi_0$ and $B_{0z}$. For angular velocity
profile~(\ref{omega}) the value of $\Phi_0$ is uniquely related
to the value of $\Omega_1=\Omega(R_1)$. The velocity of the 
plasma should be subsonic everywhere for 
Boussinesq approximation to be a reasonable approximation. 
At supersonic speeds, plasma is pushed close to the outer wall
leaving very rarefied region in the bulk between cylinders.
In this case, it is more appropriate to discuss buoyant or Parker
instabilities than MRI.
The velocity of rotation is maximum near the inner cylinder,
therefore, our calculations are valid for $\Omega_1 R_1 < C_s$.
The maximum value $\Omega_{1{\rm max}}=C_s/R_1$ is given in 
Table~\ref{table1}.

We plotted the MRI growth rate $\mbox{Im}\,(\omega)/\Omega_1$ 
for $m=0$, $k_z=\pi/L$ mode in Fig.~\ref{fig1}. 
We also searched for the unstable non-axisymmetric modes, but
did not find any. In a small region to the right in Fig.~\ref{fig1} 
with small contour separation the mode is purely growing ($\mbox{Re}
\,(\omega) =0$), in the rest of the unstable region the mode is 
growing and oscillating ($\mbox{Re}\,(\omega) > 0$).  
The MRI is absent for weak magnetic fields ($v_{{\rm A}z}/C_s \ll 10^{-2}$) 
and small wave numbers ($k_z L < 2\pi$) because driving force from 
magnetic field is too weak to overcome the viscosity and diffusivity. 
The MRI is also suppressed in high magnetic field region due to 
magnetic tension. 
For a given magnetic field, only 
finite number of modes in the $k_z$ direction will be excited.

The drawback of using plasmas is limited time available
before plasma recombines enough to significantly reduce its
coupling with the magnetic field. 
During the confinement time of the order of $10^{-3}\,\mbox{s}$ 
initial perturbations
grow by $\sim 10$ times, and can be 
registered experimentally.
This confinment time is within the range of planned experiment.

\begin{theacknowledgments}
The authors are grateful to Zhehui Wang of Los Alamos National Laboratory
for encouraging the present work and support. Discussions with Hantao Ji
and Jeremy Goodman were very useful. 
KN acknowledges support from DOE through the LDRD-ER program at
Los Alamos National Laboratory.
VP acknowledges support from 
DOE grant DE-FG02-00ER54600.
\end{theacknowledgments}

\end{document}